# Three Levels of Modeling: Static (Structure/Trajectories of Flow), Dynamic (Events) and Behavioral (Chronology of Events)

Sabah Al-Fedaghi
Computer Engineering Department
Kuwait University
Kuwait

*Abstract*—Constructing a conceptual model as an abstract representation of a portion of the real world involves capturing the (1) static (things/objects and trajectories of flow), (2) the dynamic (event identification), and (3) the behavior (e.g., acceptable chronology of events) of the modeled system. This paper focuses on examining the behavior notion in modeling and current works in the "behavior space" to illustrate that the problem of behavior and its related concepts in modeling lacks a clear-cut systematic basis. The purpose is to advance the understanding of system behavior to avoid ambiguity-related problems in system specification. It is proposed to base the notion of behavior on a new conceptual model, called the thinging machine, which is a tool for modeling that establishes three levels of representation: (1) a static structural description that is constructed upon the flow of things in five generic operations (activities; i.e., create, process, release, transfer and receive); (2) a dynamic representation that identifies hierarchies of events based on five generic events; and (3) a chronology of events. This is shown through examples that support the thinging machine as a new methodology suitable for all three levels of specification.

## I. Introduction

The main objectives of conceptual modeling of software and systems include enhancing understanding of the modeled system, providing a point of reference for designers to assemble requirements and specifications, and documenting the system for future reference. Constricting a conceptual model as an abstract representation of a portion of the real world involves capturing (1) the static (things/objects and trajectories of flow), (2) the dynamic (events identification), and (3) the behavior (e.g., acceptable chronology of events) of the modeled system.

### A. Problem by Example

The issue is that such a three-level conceptualization is not clearly recognizable in most current modeling methodologies in software and system engineering. To exemplify such problems, consider the issue of behavior specification of flow models. According to Bock and Gruninger [1], "Flow models are the most common form of behavior specification. They underlie popular programming languages and many graphical behavior specification tools. However, their semantics is typically given in natural language or in varied implementations, leading to unexpected effects in the final system." Bock and Groninger [1] discussed a way to remove ambiguity by restating flow modeling constructs in terms of constraints on runtime sequences of behavior execution. In this context, ambiguity refers to omitting information. Bock and Gruninger [1] gave an example of this problem as shown in the activity diagram of Fig. 1 where (1) the arrow in the figure is often interpreted as signifying that the paint behavior sends a message to the dry behavior or (2) the arrow means that dry must always happen after paint whenever the paint behavior is performed [2]. Fig. 1 is actually intended to state that the execution of the ChangeColor behavior is an execution of the paint behavior, after which an execution of the dry behavior will occur [1].

In this paper, the claim is that Fig. 1 mixes up the notion of activity with the notion of event and shows the static description with the dynamic description. In the current paper, we will define "activity" and "event" and present static and dynamic models that substantiate our claim.

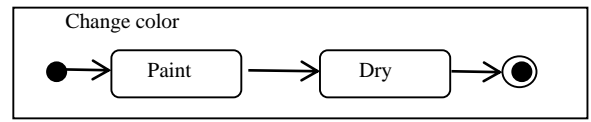

Fig 1. Behavior notation example (Adapted from [1]).

### B. Problem at Large in Modeling: What is Behavior?

Behavior has constituents that define it and form a network or a "space" of linked behavior-based concepts. According to Deleuze and Guattari [3], "Every concept has components and is defined by them [and] there is no concept with only one component." In this paper, we examine the behavior plane in modeling and review current works in the "behavior space" to illustrate that the problem of behavior and its related concepts in modeling lacks a clear-cut systematic basis. In spite of many examples given in the related literature, behavior, from our perspective, does not have a direct reply to the problem.

The issue is that "a behavioral description presumed a structural description, but a structural description also presumed a behavioral description" [4]. Moreover, the inability to connect structure and behavior is sometimes frustrating [4]. Looking at UML, "The notations provided by the UML for describing behavior are complex, poorly defined and poorly integrated, that round tripping between code and model is far too loose and error prone, and that tools in general are poor in how they integrate modelling artefacts into the lifecycle" [5].

*C. Our Contribution*

We will propose basing the notion of behavior on a new conceptual model, called the thinging machine (TM). TM is a tool for modeling, which in this context is directed at understanding, in contrast to the general notion of modeling for he purpose of prediction. We will establish three levels of representation as follows:

*1)* A static structural description that uses a single ontological element called a thimac (thing/machine). The thimac simultaneously has the structure of a thing (e.g., an object) and a machine (i.e., a process). Flow trajectories are based on five generic operations: create, process (change), release, transfer, and receive.

*2)* A dynamic representation that identifies events in the thimac to form hierarchies of events to divide the execution of the modeled system into segments of time.

*3)* A chronology of events that specifies the acceptable behavior of the system.

*D. Organization of the Paper*

The paper is organized as follows: Section 2 explores current conceptualizations of such terms as "behavior," "action," and "event" to demonstrate the arbitrary and sometimes contradictory use of these notions. To achieve a self-contained paper, in Section 3, we review TM with enhancement of its details. Section 4 illustrates TM modeling by providing a new example. Section 5 discusses the drying paint example, given in Subsection *A* of this introduction, in terms of TM. Section 6 clarifies the notions of activity and behavior as the basis of static and dynamic models, respectively. Section 7 shows that the TM diagram unifies UML activity and sequence diagrams.

## II. A Current Sample of Semantics: Behavior, Activity and Actions

This section highlights descriptions of the notion of behavior and its related concepts. Published works provide plenty of examples of behavior specifications, behavior execution, behavior taxonomies, behavior occurrences, behavior models, behavior specialization, runtime behavior, etc., but, in many references, a single statement about *what behavior is* does not exist. Accordingly, we will highlight interpretations from various works, mostly in software engineering, focusing on the terms "behavior," "actions," "activity," and "events." We will present only high points of exploration, as extensive material about the topics are scattered inside many research papers that have been published.

*A. Behavior*

In UML, a basic segmentation of "behavioral" and "structural" features exists, where "behavior is a function of time and structure is a function of space" [4]. UML is excellent at distinguishing them, "but not so good at putting them back into a meaningful relationship" [4]. According to the Object Management Group [6], "All behavior in a modeled system is ultimately caused by *actions* executed by so-called 'active' objects.'" Actions ultimately cause "all behavior in a modeled system" and "all behavior is the consequence of the actions of structural entities" [7].

Operations may be bound to activities or other behaviors [7]. A behavior describes possible executions, and an execution is the performance of an algorithm according to a set of rules [7].

Behavior specification is called "activity" and occurrence is a runtime execution of a behavior specification. In UML 2, behaviors are described as classes, and their executions are instances. For example, ChangeColor in Fig. 1 is a class, and each time it is executed, a new instance is created [1].

It is proposed (e.g., [1]) to specify the semantics of the UML activities using a process specification language (PSL). A PSL activity is said to be a reusable behavior (e.g., ChangeColor or Paint in Fig. 1) and is equivalent to the UML 2 concept called "behavior." A PSL occurrence is a runtime execution of an activity [1].

In UML, most of the time, behavior means behavioral diagrams (sequence diagrams, activity diagrams, and state machine diagrams) that "depict the elements of a system that are dependent on *time* and that convey the *dynamic concepts* of the system and how they relate to each other. The elements in these diagrams resemble the *verbs* in a natural language and the relationships that connect them typically convey the *passage of time*." [8] (Italic added for emphasis). Some UML literature refers to verbs as behavioral *things* and nouns as depicting the static behavior of a model [9]. In software design, we may "think of the linguistic analogy [about verbs and nouns]: nouns are like business objects and verbs are like use cases. When we're children, we don't start talking in verbs, we start pointing at things and saying their names" [4].

Dynamic behavior shows collaborations among objects and changes to the internal states of objects [10]. Dynamic behavior is usually defined in terms of behavior (e.g., "the dynamic behavior is the behavior of the system when it is running/operating") [10].

In philosophy, an agent's behaviors are not actions: "Actions are definitely different from the bodily movements that are controlled by non-cognitive homeostatic processes or reflexes" [11].

*B. Action*

Actions are used to define fine-grained behaviors. An action takes inputs and converts them into outputs. Basic actions include calling operations, sending signals, and invoking behavior [7].

An action represents a single step within an activity. An activity represents a behavior that is composed of actions. Examples of actions are sending and accepting a payment [7].

"The action concept is present everywhere the dynamic aspects of the world are to be taken into account. In some domains (e.g., dynamic logic), actions are confused with events" [11].

In philosophy, the concept of action is difficult to grasp [11]. Trypuz [11] lists some of these meanings of "action": an event carried out by an agent, an event caused by an agent with the intention to do this action, and an event caused by an agent for a reason.

Linguists distinguish lexical aspectual classes of verbs and verb phrases by their relation with time: activity (e.g., run or eat), state (e.g., know, be sick, or sit), accomplishment (e.g., eat an apple, or climb a mountain) and achievement (e.g., realize, reach the summit).

*C. Events*

According to [12], an "event is something that 'happens' during the course of a process. Events affect the flow of the process. Several types of event exist: TimerEvent, ConditionalEvent, etc."

An event is a stimulus that triggers state changes. Events are representations of requests from other objects. An event is defined as the specifications of noteworthy occurrence that has an allocation in time and space [13].

*D. No Systematic Ontology*

In spite of our attempt to put the conceptual highlights given above into a coherent framework, we ended by giving up such a maneuver. Instead, we opted to project the concepts over the TM model to observe their interrelatedness and connections, as shown in the remaining part of this paper.

### III. THINGING MACHINE

This section will briefly review the TM model to establish TM as a foundation to study behavior. A more elaborate discussion of TM's philosophical foundation can be found in [14-20].

The TM ontology is based on a single category called thimacs. A thimac is a categorical wrapper that embraces classical entity-ness: objects or processes. It is simultaneously an object (called a *thing*) and a process (in the broad sense) (called a *machine*)—thus, the name "thimac." The thimac notion is not new. In physics, subatomic entities must be regarded as particles and as waves to describe and explain observed phenomena [21]. According to Sfard [22], abstract notions can be conceived in two fundamentally different ways: structurally, as objects/things (static constructs), and operationally, as processes. Thus, distinguishing between form and content and between process and object is popular, but "like waves and particles, they have to be united in order to appreciate light" [23]. TM adopts the notion of duality in conceptual modeling, generalizing it beyond mathematics.

In a thimac's two modes of being, "structural conception" means seeing a notion as an entity with a recognizable internal structure and specified trajectories of motion (called "flow" in TM). The behavioral way of conceiving thimacs emphasizes the dynamic aspects in terms of events (thimacs embrace time machines). Accordingly, we can identify a chronology of events to specify the accepted behavior.

The term "thing" relies more on Heidegger's [24] notion of "things" than it does on the notion of objects. According to Heidegger [24], a thing is self-sustained, self-supporting, or independent—something that stands on its own. A thing "things"; that is, it gathers, unites, or ties together its constituents in the same way that a bridge unifies environmental aspects (e.g., a stream, its banks, and the surrounding landscape).

The term "machine" refers to a special abstract machine called a "thinging machine" (see Fig. 2) that encapsulates the laws of flows. TM is built under the postulation that only five generic actions/operations are performed on things: creating, processing (in the sense of changing), releasing, transferring, and receiving.

A thimac (a simple or complex form of TM) has dual being as a thing and as a machine. A thing is defined as that which is created, processed, released, transferred, and/or received. A machine is defined as that which creates, processes, releases, transfers, and/or receives things. Since a thimac is a thing and a machine at the same time, we will alternate between the terms "thimac," "thing," and "machine" according to the context.

The five TM flow operations (also called stages) form the foundation for thimacs. Among the five stages, the flow (a solid arrow in Fig. 2) of a thing means the trajectory of a thing's "motion," which occupies different stages. The arrow represents a projected flow just as, say, the path of the Nile on a map.

The TM diagram reflects the succession that is imposed on this "motion" of the thing: create→release→transfer, etc. The flow among the five stages is the law of flow though the thimac. The flow is the occupation of different stages at different times. In TM, a thing has no other place to be besides the five generic stages. Note that this definition is inspired by Russell's definition of motion as occupying different places at different times [25]. Adopting this theory (used to solve Zeno's paradoxes [25]), the arrows in Fig. 2 have no corresponding events (times), as they do not denote transitions.

The generic TM flow operations can be described as follows:

- *Arrival*: A thing occupies the first stage (input gate) of a new machine.

- *Acceptance*: A thing is permitted to occupy the accept stage in the machine. If arriving things are always accepted, then arrival and acceptance can be combined to become the "receive" stage. For simplicity, this paper's examples assume a receive stage.

- *Release*: A thing occupies a release stage where it is marked as ready to be transferred outside of the machine.

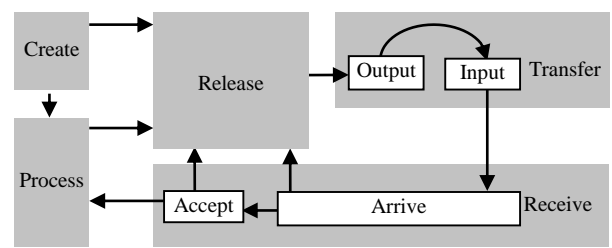

Fig 2. A Thinging Machine.

- *Transference*: A thing occupies a transfer stage (output gate) to be transported somewhere outside of the machine.

- *Creation*: A new thing is born (created) in a machine. A machine creates, in the sense that it "finds/originates" a thing; it brings a thing into the system and then becomes aware of it. "Creation" can designate "bringing into existence" in the system because what exists is what is found.

In addition, the TM model includes memory that is accessed from all stages and triggering (represented as dashed arrows) that connects thimacs in non-flow ways (e.g., classical control flow among independent programs that have no data flow among them).

IV. THINGING MACHINE BY EXAMPLE

To illustrate TM, we use the script model proposed by Schank and Abelson [26] that represents people's knowledge of events in terms of stereotyped sequences of routine actions. Fig. 3 is an example of the script for "going to a restaurant" [26] that describes the sequence of actions happening in a restaurant. In Fig. 3, the preconditions are indicated by the entry conditions "customer is hungry" and "customer has money," and consequences are marked by the results "customer has less money," "customer is not hungry," and "owner has more money." The script is divided into scenes (e.g., entering, ordering, eating, and exiting) and actions that fall under various scenes. According to Chen [27], "A series of psychological studies indicates that *scripts* correspond to psychological reality, in the sense that people indeed use predetermined, stereotyped structures to understand routine events and that people have significant agreement on the actions that comprise these events."

*A. Static TM Model*

The TM model of such a script is shown in Fig. 4 and can be explained as follows.

- First, the customer (Circle 1) flows to the restaurant (2). Note that the customer thimac contains two subthimacs: the state of being hungry (3) and the money machine that he or she has.

- Upon entering the restaurant, the customer activates (triggers; 5) looking around (6) that triggers a decision about where to sit (7). The decision triggers moving (8) to a table (9).

- Next (this sequence will be specified in the TM dynamic model), the customer takes the menu (10 and 11) and processes it (12) to trigger ordering food (13).

- The food order flows (14 and 15) to the waiter (16), who takes it (17 and 18) to the cook (19).

- The cook creates the food (20) and gives it (22) to the waiter, who receives the food (23) and carries it (24) to the customer (24).

- The customer eats the food (25).

- When the customer finishes eating, the waiter gives the customer the bill (26 and 27) and leaves the table (28 and 29).

- Then, the customer leaves the table (30) and goes (31) to the cash register (32), where he or she pays (show; 33) money (34) that flows to the cash register (35).

- The customer leaves the restaurant (37) in a state of being full (27) with less money (38).

The static TM model is static because it is a conceptualization that includes all trajectories of flow according to TM. The TM enforces order on the flow in a thimac. The static model is just the mental memory: everything is there, now, existing in the same memory. If there is a flow from X to Y and a flow from Y to X (e.g., traffic on a one-lane street), then both flows are in the static state, despite the apparent contradiction that will be resolved when time is taken into consideration. It is important to note that the sequence of stages of flow will have some influence (not all) on the sequence of events, because the logical flow inside TM cannot be violated, as will be described next.

---

**Script**: Restaurant
**Entry Conditions**: Customer is hungry.
Customer has money.
Scenes:
　**1. Entering**
Customer goes into restaurant. ($E_1$)
Customer looks around. ($E_2$)
Customer decides where to sit. ($E_3$)
Customer goes to a table and sits down. ($E_4$)
　**2. Ordering**
Customer picks up a menu. ($E_5$)
Customer decides on food. ($E_6$)
Customer orders food from waiter. ($E_7$)
Waiter tells cook the order. ($E_8$)
Cook prepares food. ($E_9$)
　**3. Eating**
Cook gives food to waiter. ($E_{10}$)
Waiter gives food to customer. ($E_{11}$)
Customer eats food. ($E_{12}$)
　**4. Exiting**
Waiter writes out check. ($E_{13}$)
Waiter brings check to customer. ($E_{14}$)
Customer gives tip to waiter. ($E_{15}$)
Customer goes to cash register. ($E_{16}$)
Customer gives money to cashier. ($E_{17}$)
Customer leaves restaurant. ($E_{18}$)

Fig 3.　The Restaurant Script. Adapted from [26].

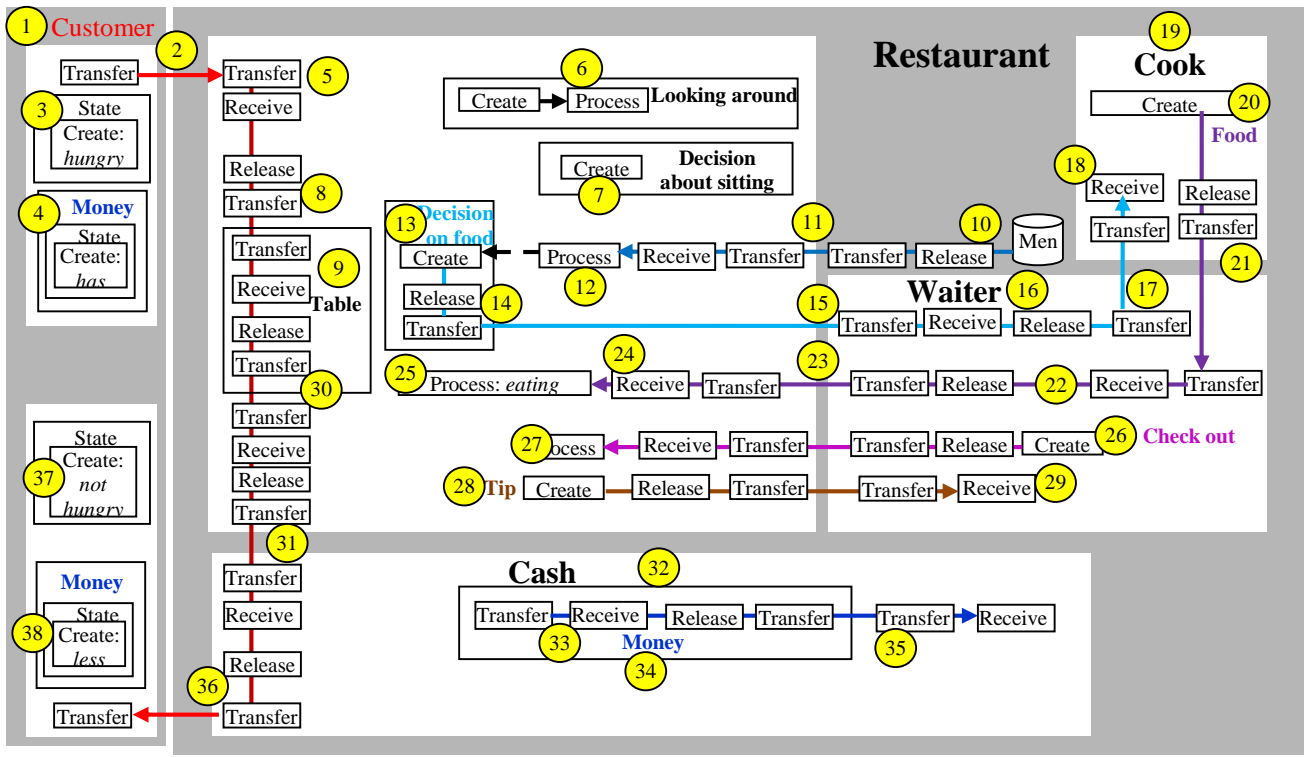

Fig 4. The TM Model of the Script.

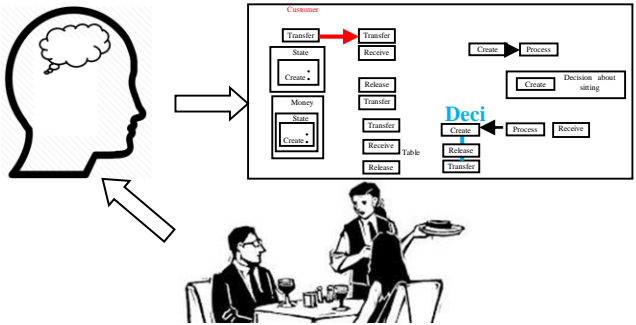

Fig 5. Static model is an encoding of a mental construct.

### B. TM Static Model and Time

In this section, we explain what is meant by calling Fig. 4 a static model. For us, Fig. 4 is a conceptual model, in the sense that it is a mental construct represented in a diagram, as shown in Fig. 5. Inside the static model, activities (operation/stages of TM or sequences of these operations) are not events (i.e., not happening in time). The activities (series of operations/stages) in the diagram and their successions are a logical progression enforced by TM, which acts as the law of flows. We note that Fig. 4 has several "floating" (cut from each other) subdiagrams with no indication of sequencing. However, the succession of stages inside a TM is compulsory. Triggering may be added for clarity, e.g., cause-and-effect.

It is possible to create the model with no consideration of succession except logical sequencing. For example, imagine that a designer captures each scene in the script on a different day. The first day he or she asks the restaurant manager to show him or her the scene of ordering, which the designer models using TM. On the second day, the designer asks to watch the paying scene, etc. At the end, he or she will end up with independent models of each scene. Then, the designer constructs Fig. 4 according to the thimac/subthimac relationship, with no idea of the ordering of the scenes: ordering, paying, etc. We say that the resultant model is a static description, because the time succession is not taken into consideration, except for the logical succession of the TM's operations: create, process, release, transfer, and receive. In modeling, we specify, in the static description, the entities and their flows, then, in the dynamic model, we identify the events in preparation for specifying the total behavior of the system.

To develop the notion of a dynamic TM model, we need the notion of an event. An event in TM is a thimac that includes a time. For example, the event *the customer goes into restaurant* is modeled as shown in Fig. 6. It includes the time, the region where the event occurs, the event, and other thimacs (e.g., intensity) that are not shown in this example.

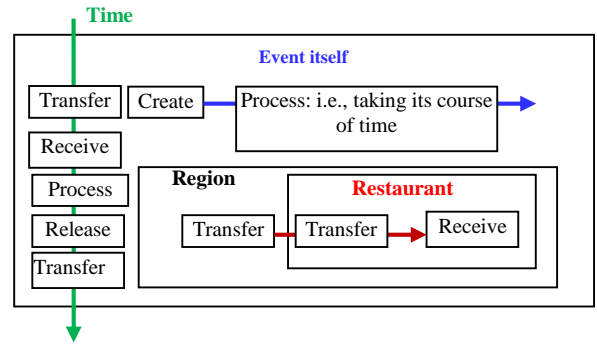

Fig 6. The TM Model of the Event "Customer attends Restaurant."

**Identifying events based on the five generic operations**

Generic events correspond to generic TM operations (e.g., receive is viewed at the dynamic level as the receive *event* with time.) We said previously that TM encapsulates the laws of flow. Flows decide the chronology of generic events. For example, the receive event occurs after the transfer event. We can build events of events, thus forming hierarchies of events based on the five generic TM operations

**Identifying events among disconnected thimacs**

Additionally, the behavior of the system requires "linking" the "floating" (cut from each other) subdiagrams, some of which are connected by triggering, discussed previously. The links are non-TM links (roughly corresponding to the so-called "control flow") and decided according to some type of cause-and-effect observation. In the given restaurant example, looking around happens after entering and the decision to sit happens after looking around.

In current modeling methodology (e.g., UML), non-TM specifications are mixed with the "data flow" from the beginning of writing the specification. For example, UML activity diagrams include data flow and control flow simultaneously. Hence, in the data flow parts, the behavior is decided by the data flow. In the control flow part, the behavior has one stream of succession decided at the beginning of the specification, in a way that is similar to the procedural programs method. Thus, in the restaurant example, we see that "entering," which includes a sequence of activities, is followed by "looking around," which also includes many sequential activities inside it, then "deciding where to sit," etc. It functions just like a main program with a set of subprograms. It is possible to permit a system behavior where the customer sits immediately after entering. However, such a chronology of events is not permitted in the given script.

Confusion exists between operations (activities, i.e., sequence of operations) and events. Activities, such as send and receive (e.g., sequence diagram), are viewed as events (i.e., operation plus time). However, this is correct only if we specify a single chronology of events. Obviously this is very restrictive modeling. Imagine a person's behavior is specified only as wake up→eat→work→home→sleep. The TM model of behavior permits specifying all other sequences that form acceptable behaviors. Current methods of modeling overcome this restriction by specifying each behavior by itself, as will be shown in the next example.

We observe that this single thread of behavior is the cause of mixing up activities and events as appears in the so-called behavior diagrams. In TM, the dynamic model is developed after finishing the static model. As we see in the restaurant script in Fig. 3 (a type of activity diagram), entering is an activity and an event, looking around is an activity and an event, etc. So the behavior becomes the chronology of activities instead of the chronology of events. The result is a single behavior: entering→looking around→making decision →sitting, etc. An activity in TM is a generic operation or a series of generic operations that works correctly as a chronology of events, as long as the series of TM operations continues (e.g., in data flow). However, this does not work for multiple acceptable behaviors if different types of flow exist.

The static TM model induces flow that partially constrains the behavior (chronology of events). Additionally, we have to weave the "floating" (cut from each other) subdiagrams into different streams of events to specify permitted behavior. In the restaurant example, looking around *may* happen (be permitted) after sittin*g* (e.g., a regular customer may go straight to a preferred table without looking around for best seat) or vice versa (looking around to signal a waiter). The dynamic model specifies events at a certain level or above the generic events. The chronology of events specifies the legal behavior of the system.

In the restaurant example, for simplicity's sake, we represent each event by its region. Accordingly, each step taken in the scene in Fig. 4 is an event. Fig. 7 shows the events of the script and Fig. 8 shows the behavior of the script system in terms of the chronology of events as given.

However, the dynamic TM model makes it possible to specify other types of behavior in the behavior specification (chronology of events). For example, suppose that we permit the following two types of behavior:

- An old customer with a favorite table enters and sits down without looking around and deciding where to sit.
- The chef cooks the food before customers enter the restaurant.

Fig. 9 shows the new chronology of events where the behavior can starts at event 1 or event 9.

## V. TM Model of the Paint–Dry Example

In this section, we show that it is not good to regard an activity (e.g., Paint and Dry in Fig. 1) as a behavior. In this type of modeling, behavior is an action (a TM stage or series of stages). In TM, Paint→Dry refers to the flow from the Paint (-ed) compass thimac to the Dry thimac. Because it is a flow, the static model enforces a sequence of TM stages.

According to Bock and Odell [28], occurrences (of "the things being modeled") are supposed to obey models of behaviors. Apparently, their use of the term *behavior* (as in an activity diagram) corresponds to the static TM model. A static diagram models behavior in a superficial way, based on the ambiguous notions of data flow and control flow. The result limits the specification of multiple behaviors.

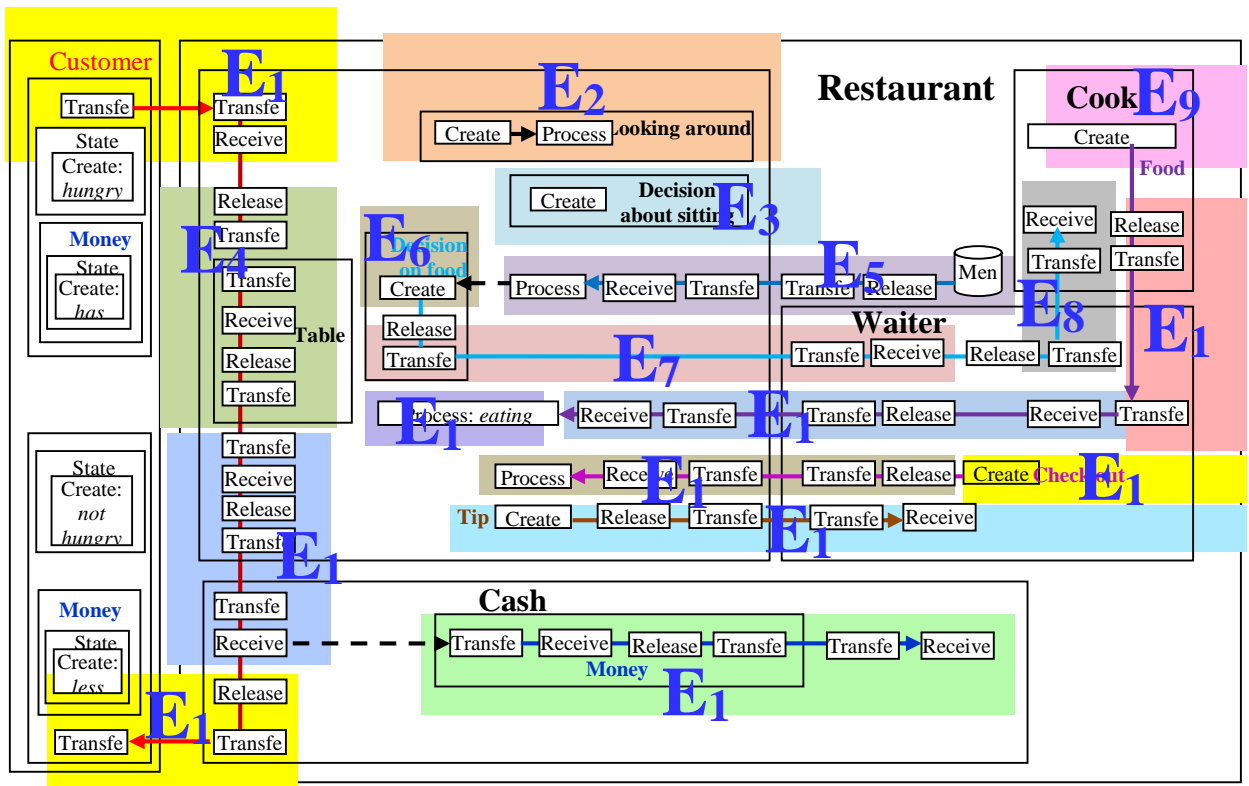

Fig 7. The Events in the TM Model of the Script.

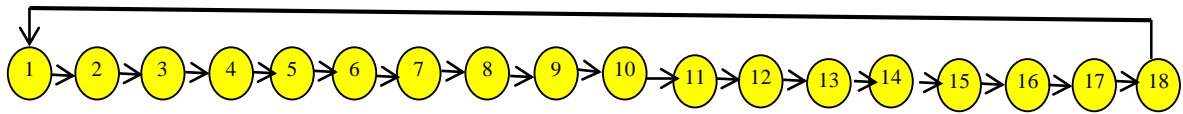

Fig 8. The Chronology of Events in the Script.

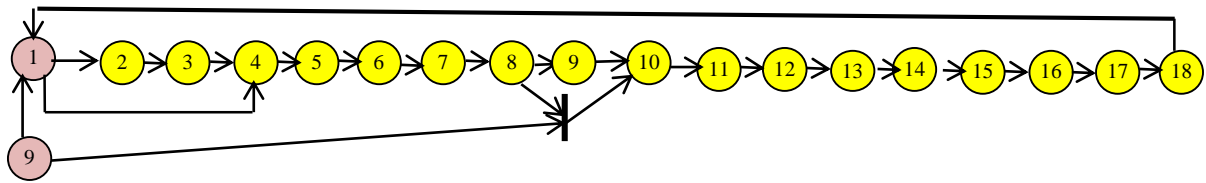

Fig 9. The New Chronology of Events.

Bock and Odell [28] stated that UML has three ways of specifying behaviors: activities (Fig. 1), state machines, and interactions. In this context, "UML behavioral diagrams depict the elements of a system that are dependent on time and that convey the dynamic concepts of the system and how they relate to each other. The elements in these diagrams resemble the *verbs* in a natural language and the relationships that connect them typically convey the passage of time. For example, a behavioral diagram of a vehicle reservation system might contain elements such as *Make a Reservation, Rent a Car, and Provide Credit Card Details*." [8] (Italic added).

From the TM point of view, in Fig. 1 ChangeColor, Paint, and Dry are all thimacs in TM. Fig. 10 shows the TM representation of Fig. 1.

In the figure, the color (material) flows from its place (e.g., a can; Circle 1) to the compass (2) to be processed (painted; 3) and then allowed to dry (4). The sequence (succession, following after) of stages release→transfer, etc., is a logical sequence (in agreement with the structure of TM) that may or may not coincide with the time sequence (in this case, it does). Flows in the static TM model "exist" (appear) simultaneously and all "exist" in the static "world" together "now" as in maps. In general, as we saw in the restaurant example, it does not take time into consideration. The static model embeds the union of all behaviors.

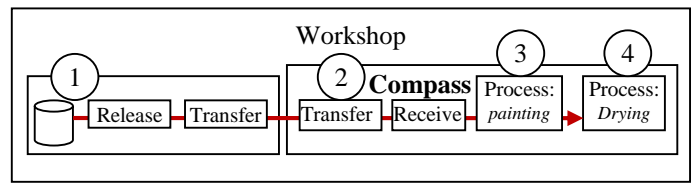

Fig 10. The Static TM representation of the Overlapping Behaviors.

Suppose that the events are selected as shown in Fig. 11, then two possible behaviors are shown in Fig. 12: $E_1 \rightarrow E_2$ and $E_1 \rightarrow E_3$. The static description embeds all possible events and flows do not represent a specific behavior (a series of events). The static model (Fig. 10) only shows the operation (activities), not the behavior. The operations (activities) can be any of the five generic TM operations or a series of them. The static model does not involve events; however, all potential behaviors are "sleeping" together through their operations. We distinguish (based on TM) events in the dynamic model.

We can see the origin of the ambiguity in Fig. 1, which is a type of static description, because it includes operations (activities), not events. The time factor ambiguously appears as a by-product, because of the arrows. Thus, in general, we differentiate between logical succession of operations and time succession, hence between thimacs without time and thimacs with time. As a consequence, the behavior is not attached to the static TM model but it appears with the dynamic TM model that embeds time.

This conclusion will be clarified further with a more complex example in the next section.

## VI. WHAT EXISTS IS NOT NECESSARILY WHAT HAPPENS

Behavior is typically defined as movement, activity, or process. However, behavior is not necessarily a movement (e.g., adult male mountain gorillas need not move to emit fear scents and octopuses often rest motionless when camouflaging against predators) [29]. Moreover, behavior is not change or activity (e.g., sweating). A process specifies behavior only if it includes events. A process consists of a system of interdependent relations between the objects, events, and other entities in an environment [29]. Events imply embedding in time. The origin of mixing these notions is the initial ontological (object-oriented or process-oriented) assumption where *what happens* (e.g., an event) and *what exists* (e.g., sticks and stones) are differentiated. In TM, the thimac *is there* and, simultaneously, the thimac *happens*. The event is a thimac that *happens* and the thing (object) is the thimac *there*.

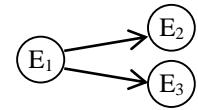

Fig 12. The Behavior of Paint→Dry.

The static thimac description is a thing and its behavior is based on its time-version that specified the chronology of its subevents. Calling a static description (e.g., activity, state and interaction diagrams in UML) "behavior" is unfortunate. The order in such a description is a transaction-based (i.e., flow-based) order not a time-based order.

### A. Example

Bock and Gruninger [2] specify that a food service process must include ordering, preparing, serving, eating, and paying, but not necessarily in that order. The constraints may be (1) ordering, preparing, and serving always happen before eating, (2) serving happens after preparing and ordering, and (3) paying can happen anytime in the process. Four kinds of food services are given, represented as activity diagrams, as follows [2]:

- FastFoodService: Prepare→Order→Pay→Serve→Eat
- RestaurantService: Order→Prepare→Serve→Eat→Pay
- Buffet: Prepare→Order→Serve→Pay→Eat
- ChurchSupper: Pay→Order→Prepare→Serve→Eat

Bock and Gruninger [2] construct a separate activity diagram for each service. This is a model that mixes up activities with events. FastFoodService, RestaurantService, Buffet, and ChurchSupper are different behaviors of the same system.

Fig. 13 shows the static TM representation for this system as a single diagram. We note that ordering, paying, serving, and preparing subdiagrams are "loose" in their time relationships with other subdiagrams. However, inside each of the four services, the static TM model obeys the time flow to develop the events of the system by preserving flow order and converting the five generic TM operations into generic events that follow the same sequence. In UML, basic events are specified in the sequence diagram as sending of a message and receiving of a message [30].

Fig. 14 shows selected events of the food-ordering system. Hence, we can assemble the loose subdiagrams into several forms that reflect the acceptable behavior of the system. The loose subdiagrams are events, where the interior of each of them is fixed with respect to the TM operations.

Fig. 15 shows the behavior of the system. Events may be repeated in the diagram for clarity. Fig. 16 shows the union of these events.

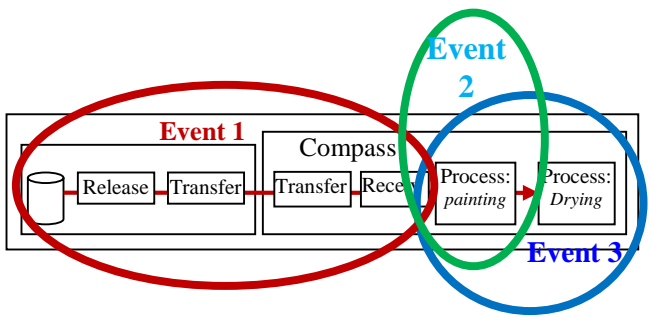

Fig 11. The Events in Paint→Dry.

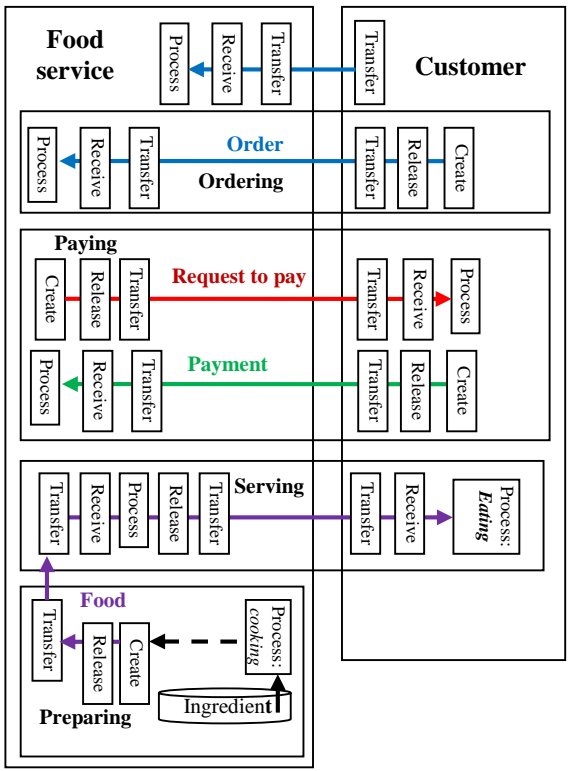

Fig 13. The Static TM Model of a Food Service System.

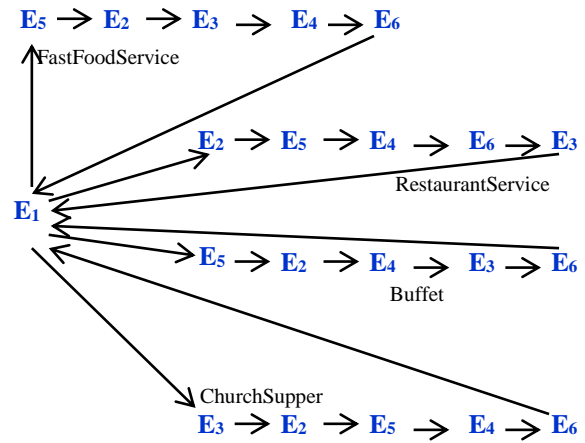

Fig 15. Behavior of the Services System.

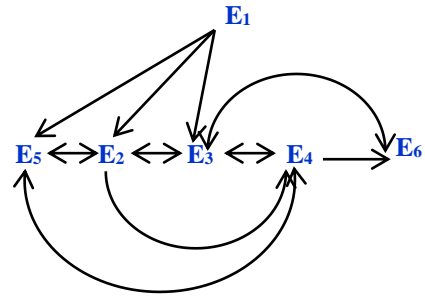

Fig 16. The Union of Behaviors of Different Services.

## VII. CONCLUSION

In this paper, we have proposed a possible approach to defining behavior in modeling, including behavior-related related concepts (e.g., operation, action, and event). Accordingly, we defined the following basic concepts:

- Behavior is the chronology of events in the TM model

- Events are thimacs with time submachines

- Operations/actions are the five generic TM operations and the series of these operations.

TM can be used as a modeling tool that forms three levels of representation with basic elements of operations/actions, events, and behavior.

Representing a model in a single diagram may be raised as an issue. In TM, the world is abstracted as thimacs with five generic stages. The grand thimac is not a single, monolithic, unmanageable whole; instead, it incorporates decomposability by its skeletal structure of multiple interior thimacs. This decomposability is based on joints (flows and triggering among thimacs) that form the structure (anatomy) of a system (the overarching thimac). Accordingly, the conceptual model is a single diagram, but the implementation (e.g., software) lends itself to differentiating thimacs at the joints via an adequate conceptualization.

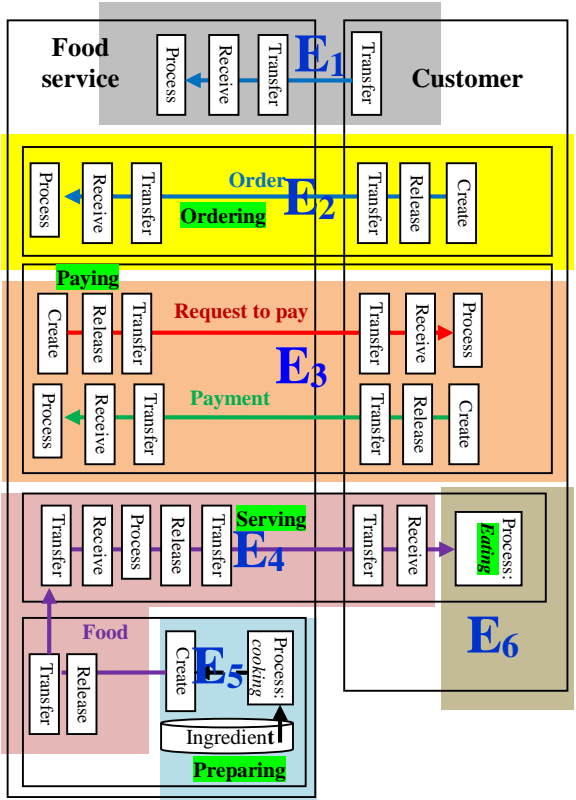

Fig 14. The Events in the Food Services System.